\documentclass{article}
\begin{document}
\hoffset = -2 truecm \voffset = -2 truecm
\title{Gap equations and effective potentials at finite temperature and chemical potential
in D dimensional four-fermion models\footnote{This work was supported by the National Natural Science Foundation of China.} \\
}
\author{Zhou Bang-Rong  \\
Department of Physics, Graduate School of the \\
Chinese Academy of Sciences, Beijing 100039, China \\
and CCAST (World Lab.),P.O. 8730, Beijing 100080, China }
\date{}
\maketitle
\begin{abstract}
We have proven the general relations between the gap equations obeyed by dynamical fermion
 mass and corresponding effective potentials at finite temperature and chemical potential
in D dimensional four-fermion interaction models. This gives an easy approach to get
effective potentials directly from the gap equations. We find out explicit expressions for
the effective potentials at zero temperature in the cases of $D=2,3$ and 4 for practical use.
\end{abstract}
\indent Four-fermion interaction models \cite{kn:1,kn:2} are good
laboratories to research phase transitions at finite temperature and finite density \cite{kn:3,kn:4}. In such research, a fundamental tool is effective potential\cite{kn:5,kn:6}, but in some cases, the gap equation obeyed by the dynamical fermion mass as the order parameter of symmetry breaking is also very useful analytic means \cite{kn:7,kn:8,kn:9}. Finding out a definite relation between an effective potential and corresponding gap equation has not only theoretical significance but also practical use. This is because by such relation one can easily calculate corresponding effective potential from a gap equation, if the gap equation may be known beforehand, and avoid many lengthy formal derivations.
In this paper we will give the definite relations between effective potentials and  gap equations in a class of four-fermion interaction models. To be general, we will take the dimension of time-space to be equal to $D$. \\
\indent The Lagrangian of the models will be expressed by
\begin{equation}
{\cal L}(x)=\sum_{k=1}^N\bar{\psi}^k(x)i\gamma^{\mu}\partial_{\mu}\psi_k(x)+
              \frac{g}{2}\sum_{k=1}^N[\bar{\psi}^k(x)\psi_k(x)]^2,
\end{equation}
where $\psi_k(x)$  are the fermion fields with $N$ "color" components and $g$ is
the four-fermion coupling constant. The discussions will be made in the fermion bubble
diagram approximation which is equivalent to the leading order of the $1/N$ expansion.

The Lagrangian (1) is equivalent to
\begin{equation}
{\cal L}_{\sigma}(x)=
\sum_{k=1}^N\bar{\psi}^k(x)i\gamma^{\mu}\partial_{\mu}\psi_k(x)
-\sigma(x)\sum_{k=1}^N\bar{\psi}^k(x)\psi_k(x) -\frac{1}{2g}\sigma^2(x),
\end{equation}
where $\sigma(x)$ is an auxiliary scalar field. From Eq.(2) we can calculate the
effective potential at temperature $T=0$  and chemical potential $\mu=0$  by standard method and obtain \cite{kn:10}
\begin{eqnarray}
V_{eff}^{(D)0}(\sigma_c)&=&\frac{\sigma_c^2}{2Ng}-\frac{1}{N}\sum_{n=1}^{\infty}\frac{\sigma_c^{2n}}{(2n)!}\Gamma^{(2n)}(l_i=0) \nonumber \\
&=&\frac{\sigma_c^2}{2Ng}+i2^{\eta_D-1}\int \frac{d^Dl}{(2\pi)^D}\ln\left(1-\frac{\sigma_c^2}{l^2+i\varepsilon}\right),
\end{eqnarray}
where $\sigma_c$ is the constant "classical field" configuration, $l_i$ denotes the external momentum of the one-particle irreducible (1PI) Green functions $\Gamma^{(2n)}$, $2^{\eta_D}$ is the trace of the spin unit matrix in $D$ dimensional time-space with
\begin{equation}
\eta_D=\left\{\matrix{D/2, &{\rm when} & D={\rm even} \cr
                      (D-1)/2, &{\rm when} & D={\rm odd} \cr}\right..
\end{equation}
From Eq.(3) we can obtain
\begin{equation}
\frac{\partial V_{eff}^{(D)0}(\sigma_c)}{\partial  \sigma_c}=
\frac{\sigma_c}{gN}-2^{\eta_D}
\int \frac{d^Dl}{(2\pi)^D}\frac{i\sigma_c}{l^2-\sigma_c^2+i\varepsilon}.
\end{equation}
In present model, the dynamical fermion mass at $T=\mu=0$ can be identified with
$m(0)=\sigma_c$, thus Eq. (5) may be changed into
\begin{eqnarray}
\frac{\partial V_{eff}^{(D)0}[m(0)]}{\partial  m(0)}&=&
\frac{1}{gN}\left[m(0)
-gN{\rm tr}\int \frac{d^Dl}{(2\pi)^D}\frac{i}{\not{l}-m(0)+i\varepsilon}
\right] \\
&=&m(0)\left[\frac{1}{gN}-2^{\eta_D}\int \frac{d^Dl}{(2\pi)^D}\frac{i}{l^2-m^2(0)+i\varepsilon}\right].
\end{eqnarray}
Noting that the second term in the square brackets in the right-handed side of Eq. (6) simply represents the tadpole diagram with one fermion loop and the whole expression in the square brackets being set to be zero or say
$$\partial V_{eff}^{(D)0}[m(0)]/\partial  m(0)=0,$$
up to a constant factor, is just the Schwinger-Dyson equation obeyed by the dynamical fermion mass $m(0)$. Assuming that spontaneous symmetry breaking occurs at $T=\mu=0$, then from $\partial  V_{eff}^{(D)0}[m(0)]/\partial  m(0)=0$
and Eq. (7), we may affirm that the gap
equation
\begin{equation}
\frac{1}{gN}=2^{\eta_D}\int \frac{d^Dl}{(2\pi)^D}\frac{i}{l^2-m^2(0)+i\varepsilon}
\end{equation}
must have solution with $m(0)\neq 0$.
Eq. (6) may be generalized to the case of finite $T$ and $\mu$ by the
replacements
\begin{equation}
m(0) \rightarrow m\equiv m(T,\mu),
\end{equation}
where $m$ is the dynamical fermion mass at finite $T$ and $\mu$, and in the real-time
formalism of thermal field theory \cite{kn:11},
\begin{equation}
\frac{i}{\not{l}-m(0)+i\varepsilon}\rightarrow \frac{i}{\not{l}-m+i\varepsilon}
-2\pi(\not{l}+m)
\delta(l^2-m^2)\sin^2\theta(l^0,\mu)
\end{equation}
with
\begin{equation}
\sin^2(l^0,\mu)=\frac{\theta(l^0)}{\exp[\beta(l^0-\mu)]+1}+
                      \frac{\theta(-l^0)}{\exp[\beta(-l^0+\mu)]+1}.
\end{equation}
Thus the effective potential $V_{eff}^{(D)}(T,\mu,m)$ at finite $T$ and $\mu$ satisfies the equation
\begin{equation}
\frac{\partial V_{eff}^{(D)}(T,\mu,m)}{\partial  m}= \frac{1}{gN}\left\{m -gN{\rm
tr}\int \frac{d^Dl}{(2\pi)^D}\left[\frac{i}{\not{l}-m+i\varepsilon} -2\pi(\not{l}+m)
\delta(l^2-m^2)\sin^2\theta(l^0,\mu) \right] \right\}
\end{equation}
It is seen from Eq.(12) that $\partial V_{eff}^{(D)}(T,\mu,m)/\partial  m=0$ is just the Schwinger-Dyson equation obeyed by $m$. After Eq. (8) is substituted, i.e. in the condition that spontaneous symmetry breaking at $T=\mu=0$ occurs, we obtain from Eq.(12) that
\begin{eqnarray}
\frac{\partial V_{eff}^{(D)}(T,\mu,m)}{\partial  m}&=&
m\left\{2^{\eta_D}\int\frac{d^Dl}{(2\pi)^D}\left[\frac{i}{l^2-m^2(0)+i\varepsilon}-\frac{i}{l^2-m^2+i\varepsilon}\right]\right. \nonumber \\
&+&\left.\frac{2^{\eta_D}}{(4\pi)^{(D-1)/2}}\frac{\Gamma(D-1)}{\Gamma(\frac{D-1}{2})}T^{D-2}\left[I_{D-1}(y,-r)+I_{D-1}(y,r)\right]\right\}, \nonumber \\
\end{eqnarray}
where we have used the denotations \cite{kn:12}
\begin{equation}
I_n(y,\mp r)=\frac{1}{\Gamma(n)}\int_{0}^{\infty}\frac{dx}{\sqrt{x^2+y^2}}
\frac{x^{n-1}}{\exp(\sqrt{x^2+y^2}\mp r)+1},
\end{equation}
with $y=m/T$, $r=\mu/T$. It is easy to verify that $\partial V_{eff}^{(D)}(T,\mu,m)/m\partial  m=0$ will give the gap equation at finite $T$ and $\mu$ after the gap equation at $T=\mu=0$ is substituted. In fact, we may obtain
\begin{eqnarray}
\frac{\partial V_{eff}^{(D)}(T,\mu,m)}{\partial  m}&=&
\frac{m}{\pi}\left[\ln\frac{m}{m(0)}+F_1(T,\mu,m)\right], {\rm when} \ D=2 \\
&=&\frac{m}{2\pi}\left[m-m(0)+F_2(T,\mu,m)\right], {\rm when} \ D=3 \\
&=&\frac{m}{2\pi^2}\left\{\frac{m^2}{2}\ln\left(\frac{\Lambda^2}{m^2}+1\right)-\frac{m^2(0)}{2}
\ln\left[\frac{\Lambda^2}{m^2(0)}+1\right]+F_3(T,\mu,m)\right\},\nonumber \\&&{\rm
when} \ D=4,
\end{eqnarray}
where
\begin{equation}
F_1(T,\mu,m)=I_1(y,-r)+I_1(y,r)
\end{equation}
\begin{equation}
F_2(T,\mu,m)= T\left\{\ln \left[1+e^{-(m-\mu)/T}\right]+
                  (-\mu\to \mu)\right\},
\end{equation}
\begin{equation}
F_3(T,\mu,m)=4T^2[I_3(y,-r)+I_3(y,r)]
\end{equation}
and $\Lambda$ in Eq.(17) is the four-fermion Euclidean momentum cut-off in the zero temperature fermion loop integrals. We see that taking the right-handed sides of Eqs.(15),(16) and (17) to be equal to zeroes will lead to the equation $m=0$ and the gap equations of the models respectively in $D=2,3$ and 4 cases which were derived in Refs. \cite{kn:9}, \cite{kn:8} and \cite{kn:7}. Eq.(13) completely determines $\partial V_{eff}^{(D)}(T,\mu,m)/\partial  m$, thus we can derive the effective potential $ V_{eff}^{(D)}(T,\mu,m)$ at finite $T$ and $\mu$ through integration over $m$, i.e.
\begin{equation}
V_{eff}^{(D)}(T,\mu,m)=\int_0^{m}dm'
\frac{\partial V_{eff}^{(D)}(T,\mu,m')}{\partial  m'},
\end{equation}
where we have selected the value of $V_{eff}^{(D)}(T,\mu,m)$ at $m=0$ to be equal to zero. We indicate that the gap equation multiplied by $m$ can separately come from the Schwinger-Dyson equation obeyed by the dynamical fermion mass without necessity of using an effective potential beforehand and the integrand in Eq.(21), as was stated above, just relates to the gap equation multiplied by m (up to a constant factor), so Eq. (21) may actually provide us an easy way to get the effective potential simply from the corresponding gap equation multiplied by $m$  and saves lengthy derivation in conventional approach. The effective potential so obtained differs from the exact one derived by Eqs. (13) and (21) at most by a overall constant factor.  However, such difference does not effect the essential conclusion reached of phase transition feature of the system. \\
\indent As several examples of applying Eq. (21), we first consider the case of $D=2$.
From Eqs. (15) and (21) we obtain
\begin{eqnarray}
V_{eff}^{(2)}(T,\mu,m)&=&\int_0^mdm'\frac{m'}{\pi}\left[\ln\frac{m'}{m(0)}+
F_1(T,\mu,m')\right]\nonumber \\
&=&\frac{m^2}{2\pi}\left[\ln\frac{m}{m(0)}-\frac{1}{2}\right]
-\frac{1}{\beta\pi}\int_0^{\infty}dk\left[\ln\frac{1+e^{-\beta(\sqrt{k^2+m^2}-\mu)}}{1+e^{-\beta(k-\mu)}}+(-\mu\to
\mu)\right].
\end{eqnarray}
This result is completely consistent with Eq.(6.3) in Ref.\cite{kn:6}. Taking the limit $T\to 0$ in Eq.(22) we will get the effective potential at $T=0$
\begin{eqnarray}
V_{eff}^{(2)}(T=0,\mu,m)&=&\frac{m^2}{2\pi}\left[\ln\frac{m}{m(0)}-\frac{1}{2}\right]
+\frac{\mu^2}{2\pi}
+\frac{1}{2\pi}\theta(\mu-m)\left(m^2\ln\frac{\mu+\sqrt{\mu^2-m^2}}{m}
-\mu\sqrt{\mu^2-m^2}\right). \nonumber \\
\end{eqnarray}
Similar discussions can be made in the cases of $D=3$ and $D=4$. We will derive only the effective potentials at $T=0$. For $D=3$, the $T\to 0$ limit of Eq.(19) gives
\begin{equation}
F_2(T=0,\mu,m)=\theta(\mu-m)(\mu-m).
\end{equation}
Substituting it into the $T\to 0$ limit of Eq.(16) and using Eq. (21) we obtain
\begin{eqnarray}
V_{eff}^{(3)}(T=0,\mu,m)&=&\frac{1}{2\pi}\int_0^mdm'm'[m'-m(0)+F_2(T=0,\mu,m')]
\nonumber \\
&=&\frac{1}{2\pi}\left\{\frac{m^2}{2}[\mu\theta(\mu-m)-m(0)]+\theta(m-\mu)
\left(\frac{m^3}{3}+\frac{\mu^3}{6}\right)\right\}.
\end{eqnarray}
For $D=4$, through changing the integral variable by $z=(x^2/y^2+1)^{1/2}$, by means of Eqs. (14) and (20), we can express
\begin{equation}
F_3(T,\mu, m)=2m^2\int_1^{\infty}dz\left[\frac{\sqrt{z^2-1}}{e^{y(z-\alpha)}+1}+
(-\alpha\to \alpha)\right], \alpha=\mu/m.
\end{equation}
From Eq. (26) it is easy to find out the $T\to 0$ limit of $F_3(T,\mu,m)$
\begin{equation}
F_3(T=0,\mu,m)=\theta(\mu-m)\left[\mu\sqrt{\mu^2-m^2}-m^2\ln\left(\frac{\mu}{m}+\sqrt{\frac{\mu^2}{m^2}-1}\right)\right].
\end{equation}
Then substituting Eq. (27) into the $T\to 0$ limit of Eq.(17) and using Eq. (21), we obtain
\begin{eqnarray}
V_{eff}^{(4)}(T=0,\mu,m)&=&\frac{1}{2\pi^2}\left\{
\frac{m^4}{8}\left(\ln\frac{\Lambda^2}{m^2}+1\right)+\frac{\Lambda^2m^2}{8}
-\frac{\Lambda^4}{8}\ln\left(1+\frac{m^2}{\Lambda^2}\right)
-\frac{m^2(0)m^2}{4}\ln\left[\frac{\Lambda^2}{m^2(0)}+1\right]\right.\nonumber \\
&&\left.+\frac{\mu^4}{6}+\theta(\mu-m)\left[\frac{\mu
m^2}{4}\sqrt{\mu^2-m^2}-\frac{\mu}{6}(\mu^2-m^2)^{3/2}-\frac{m^4}{4}\ln\frac{\mu
+\sqrt{\mu^2-m^2}}{m}\right]\right\}. \nonumber \\
\end{eqnarray}
The equations (23), (25) and (28) can be used in discussions about high density phase transitions at $T=0$ in the four-fermion interaction models in $D=2,3$ and 4 dimensional time-space which will be conducted elsewhere. \\
\indent In summary, in this paper, we have proven the general relations between the gap equations obeyed by dynamical fermion mass and the corresponding effective potentials at finite temperature and chemical potential in $D$ dimensional four-fermion interaction models. This provides a convenient approach to derive the effective potentials directly from the gap equations multiplied by the dynamical fermion mass. The latter can be obtained separately from the Schwinger-Dyson equations obeyed by the dynamical fermion mass. The effective potentials at zero temperature and finite chemical potential in $D=2,3$ and 4 dimensional four-fermion models are explicitly given for practical use.


\begin{thebibliography}{99}
\bibitem{kn:1}  Y. Nambu and G. Jona-Lasinio, Phys. Rev. {\bf 122}(1961) 345;
                {\bf 124}(1961) 246.
\bibitem{kn:2}  D. J. Gross and A. Neveu, Phy. Rev. D{\bf 10} (1974) 3235.
\bibitem{kn:3}  D. A. Kirzhnits and A. D. Linde, Phys. Lett. {\bf 42B}
                (1972) 471;
                S. Weinberg, Phys. Rev. D{\bf 7} (1973) 2887;
                {\bf 9} (1974) 3357;
                L. Dolan and R. Jackiw, {\it ibid.} {\bf 9} (1974) 3320;
\bibitem{kn:4}  A. D. Linde, Rep. Prog. Phys. {\bf 42} (1979) 389;
                R. H. Brandenberger, Rev. Mod. Phys. {\bf 57} (1985) 1;
\bibitem{kn:5}  L. Jacobs, Phys. Rev. D{\bf 10} (1974) 3956;
                B. Harrington and A. Yildiz, Phys. Rev. D{\bf 11} (1975) 779;
                U. Wolff, Phys. Lett. B{\bf 157} (1985) 303;
              T. Inagaki, T. Kouno and T. Muta, Inter. J. Mod. Phys. A{\bf 10} (1995) 2241.
\bibitem{kn:6}  A. Chodos, F. Cooper, W. Mao, H. Minakata and A. Singh, Phys. Rev. D{\bf 61} (2000) 045011.
\bibitem{kn:7} B. R. Zhou, Phys. Rev. D{\bf 57} (1998) 3171;
                Commun. Theor. Phys. {\bf 32} (1999) 425.
\bibitem{kn:8} B. R. Zhou, Phys. Lett. B {\bf 444} (1998) 455.
\bibitem{kn:9} B. R. Zhou, Commun. Theor. Phys. {\bf 33} (2000) 451.
\bibitem{kn:10} For example, see M. Quiros, in {\it 1998 Summer School in High Energy Physics and Cosmology}, eds. A. Masiero, G. Senjanovi\'{c} and A. Smirnov, World Scientific, Singapore (1998), p.187 and the references therein.
\bibitem{kn:11} N. P. Landsman and Ch. G. van Weert, Phys. Rep. {\bf 145}                (1987) 141.
\bibitem{kn:12} J. I. Kapusta, {\it Finite-temperature field theory}, Cambridge University Press, England, (1989).
\end{thebibliography}
\end{document}